\documentclass[a4paper]{PoS}

\usepackage{ucs}                
\usepackage[utf8x]{inputenc}

\usepackage{amsmath}
\usepackage{amsfonts}
\usepackage{mathtools}
\usepackage{slashed}            
\usepackage{commath}            
\usepackage{dsfont}             

\usepackage{pgf}                
\usepackage{tikz}               
\usepackage{datetime}           
\usepackage{array}              
\usepackage{float}              
\usepackage{braket}             
\usepackage{relsize}            
\usepackage{placeins}           
\usepackage{booktabs}           
\usepackage{xfrac}              
\usepackage{todonotes}

\definecolor{aiphiblue}     {RGB}{0,84,159}
\definecolor{aiphibordeaux} {RGB}{161,16,53}
\definecolor{aiphiyellow}   {RGB}{255,237,0}
\definecolor{aiphigreen}    {RGB}{87,171,39}
\definecolor{aiphipetrol}   {RGB}{0,97,101}
\definecolor{aiphired}      {RGB}{204,7,30}
\definecolor{aiphiviolet}   {RGB}{97,33,88}

\usetikzlibrary{tikzmark, calc, arrows, spy}

\newcommand{\unit}[1]{\,\text{#1}}

\newcommand{\mev}{\unit{MeV}}
\newcommand{\gev}{\unit{GeV}}

\renewcommand{\epsilon}{\varepsilon}

\newcommand{\Rplus}{\protect\hspace{-.1em}\protect\raisebox{.35ex}{\smaller{\smaller\textbf{+}}}}
\newcommand{\cpp}{\mbox{C\Rplus\Rplus}}

\title{Three neutrons from Lattice QCD}

\ShortTitle{Three neutrons from Lattice QCD}

\author{\speaker{J.-L. Wynen} and E. Berkowitz and T. Luu\\
  Institute for Advanced Simulation and Nuclear Physics Institute --- Forschungszentrum J\"ulich\\
  Helmholtz-Institut f\"ur Strahlen- und Kernphysik --- Universit\"at Bonn\\
  JARA-HPC\\
  E-mail: \email{j.wynen@fz-juelich.de}, \email{e.berkowitz@fz-juelich.de}, \email{t.luu@fz-juelich.de}}

\author{A. Shindler\\
  Facility for Rare Isotope Beams --- Michigan State University\\
  E-mail: \email{shindler@frib.msu.edu}}

\author{J. Bulava\\
  CP$^3$ Origins --- Syddansk Universitet\\
  E-Mail: \email{bulava@cp3.sdu.dk}}

\abstract{We present a study on ab-initio calculations of three-neutron correlators from Lattice QCD.\@
  We extend the method of baryon blocks to systems of three spacially displaced baryons.
  This allows the measurement of three-neutron $p$-wave correlators with total spin $S=\sfrac{1}{2}$ and  $\sfrac{3}{2}$.
  In addition, we use automatic code generation that has high flexibility and allows for easy inclusion of additional channels in the future while optimizing the evaluation of contractions.
  Our measurements were performed on a newly generated $96\times 48^3$ Clover-Wilson gauge field ensemble with
 $m_{\pi} \approx 370\mev$.
  We present preliminary results of our calculations of one pion and nucleon as well as two nucleon ($2N$) and three neutron ($3n$) correlators.}

\FullConference{The 36th Annual International Symposium on Lattice Field Theory - LATTICE2018\\
  22-28 July, 2018\\
  Michigan State University, East Lansing, Michigan, USA.}

\begin{document}

\section{Introduction}

In recent years, large advances have been made in computing multi-baryon observables using lattice QCD.
Many studies have been published on two nucleon (2N) spectroscopy, an overview of which can be found in~\cite{berkowitz:2018pts}.
Additionally, calculations of s-wave three nucleon (3N) systems, such as $^3$H and $^3$He, have been performed in the past~\cite{aoki:2012tk, doi:2018bipm}.
Three neutron (3n) interactions, however, have not been addressed by lattice QCD to date because their correlators require non-trivial momentum projections due to the Pauli-exclusion principle. Further, the number of Wick contractions are substantially larger than their s-wave 3N counterparts.

An analysis of the full 3N spectrum (s-wave and beyond) is important however.
Unknown 3N forces are responsible for the largest uncertainties in calculations of neutron rich isotopes, such as the determination of the neutron dripline or equation of state of neutron stars.
Experiments are currently not able to measure 3n interaction energies directly as such systems are very hard to create.

Lattice calculations can perform spectrometry on 3n systems but suffers its own problems.
Apart form the aforementioned issues of computing 3n correlators, measurements are expected to suffer from a worse signal-to-noise problem than the smaller 2N systems.
In this work we develop the formalism and computational tools to evaluate 3n correlators on the lattice.

\section{Formalism}

We express the 3n correlator as a contraction $C_{3n}^{SS_3}(p) = \langle O_{3n}^{SS_3} \overline{O}_{3n}^{SS_3} \rangle$.
For simplicity, we restrict the discussion to vanishing total momentum and a single relative momentum $p$.
$S$ and $S_3$ denote magnitude and third component of the total spin.
The interpolating operator $O_{3n}$ can be composed out of the individual neutron operators $n$ as shown for the sink side operator (primed arguments indicate the sink):
\begin{align}
  O_{3n}^{S\,S_3}(x'_1, x'_2, x'_3) = \big(n^{\alpha'}(x'_1)\, \Gamma_{s\,s_3}^{\alpha'\beta'}\, n^{\beta'}(x'_2)\big)\, \Gamma_{S\,S_3}^{s\,s_3\gamma'}\, n^{\gamma'}(x'_3)\label{eq:3n_interp_op}\ .
\end{align}
Here $\Gamma_{s\,s_3}$ denotes a Clebsch-Gordan coefficient projecting onto the cubic irreducible representation\footnote{For zero total momentum and $s\in\{0,\, \sfrac12,\, 1\}$, we can use spin and cubic labels interchangeably.} $s,s_3$, see~\cite{basak:2005ir}.
Expressed in terms of quark operators, the neutron operators are
\begin{align}
  n^{\alpha'}(x') = \epsilon_{a'b'c'} d_{a'}^{\alpha'}(x) \big(u_{b'}^{\beta'} \Gamma_{\frac12 \frac12}^{\beta'\gamma'} d_{c'}^{\gamma'} \big).\label{eq:neutron_op}
\end{align}

\noindent
At most two neutrons can be at the same site due to the Pauli exclusion principle.
However, in order to be able to access $S_3 = \pm \sfrac{3}{2}$, we need to place all three neutrons at three distinct sources.
The sinks, too, need to be different from each other.
This implies that we can not simply project to an $s$-wave as that would put all three neutrons at the same site in momentum-space.

We therefore need to control all three baryon momenta individually.
To this end we employ the formalism described in~\cite{doi:2012xd, detmold:2012eu} which combines quarks into baryon blocks $B$ that have the quantum numbers of a single baryon at the sink.
Here, baryon blocks are defined as
\begin{align}
  \begin{split}
    B^{\alpha^{\prime}\,\alpha \beta \gamma}_{\hphantom{\alpha^\prime}\,abc}(x^{\prime}|f_1,x_1;f_2,x_2;f_3,x_3) &= \epsilon_{a^{\prime}b^{\prime}c^{\prime}} S_{a^{\prime}a}^{\alpha^{\prime}\alpha}(f_1, x^\prime \leftarrow x_{1})\\
    &\quad\times \left[S_{b^{\prime}b}^{\beta^{\prime}\beta}(f_2, x^\prime \leftarrow x_{2}) \Gamma_{\frac{1}{2}\frac{1}{2}}^{\beta^{\prime}\gamma^{\prime}} S_{c^{\prime}c}^{\gamma^{\prime}\gamma}(f_3, x^\prime \leftarrow x_{3})\right].\label{eq:bblock_def}
  \end{split}
\end{align}
$S(f, x' \leftarrow x)$ denotes a single quark propagator of flavour $f$ from source $x$ to sink $x'$.
$B$ has only one spin ($\alpha'$) and one site ($x'$) label at the sink but is completely `open' at the source.
In the interest of readability, we shorten the notation using superindices $I$ and $X$ defined via $B_I(x'|X) \equiv B^{\alpha^{\prime}\,\alpha \beta \gamma}_{\hphantom{\alpha^\prime}\,abc}(x'|f_1,x_1;f_2,x_2;f_3,x_3)$.

Expressed in terms of baryon blocks, the correlators take the form
\begin{align}
  C_{3n}^{SS_3}(\texttt{n}^2) = F_{\texttt{n}^2}\bigg( \sum_{\genfrac{}{}{0pt}{2}{IJK}{X_1 X_2 X_3}} T_{IJK}^{SS_3}(X_1,X_2,X_3)\; B_I(x_1'|X_1) B_J(x_2'|X_2) B_K(x_3'|X_3)\bigg).\label{eq:bblock_corr}
\end{align}
The tensor $T$ encodes the combination of $\Gamma$'s and $\epsilon$'s from~(\ref{eq:3n_interp_op}) and~(\ref{eq:neutron_op}) and projects onto total spin $S, S_3$.
Note that all three baryon blocks have distinct parameters, mirrored by $T$. We note that it is not possible to separate the dependencies on flavours and sites $X$ from spins and colours $I,J,K$.

In equation~(\ref{eq:bblock_corr}) $\texttt{n}^2$ denotes a momentum shell on a cubic lattice rather than the continuum momentum $p$ from before.
Setting the centre-of-mass momentum to zero and picking a relative momentum is not sufficient to specify the full momentum projection $F$ for a three particle system.
There is some freedom in choosing $F$ but not every projection has good overlap with the physical $3n$ system.
A choice that is easy to implement and fast to compute is
\begin{align}
  F: (x'_1, x'_2, x'_3) \mapsto (+p, -p, 0),\label{eq:mom_project}
\end{align}
i.e.\ the di-neutron in eq.~\eqref{eq:3n_interp_op} is projected to zero momentum with its constituents having back to back momenta and the singled out neutron has zero momentum on its own.
The full calculation\footnote{Unfortunately, as with most calculations of this type, it is too long to present here. See in particular Section~\ref{sec:implementation}.} of the projection onto the cubic $\texttt{n}^2$ shows that this form removes some contributions from the correlator; more specifically, the di-neutron spin in~\eqref{eq:3n_interp_op} is restricted to $s=s_3=1$.

\section{Implementation / Code Generation}\label{sec:implementation}

Naively, a correlator of the form~(\ref{eq:bblock_corr}) has $N_u!\,N_d! = 3!\,6! = 4320$ different combinations of flavours and sites. 
In practice, some terms cancel and this number is slightly lower but still too large to handle by hand.
We therefore developed a software suite that generates high performance \cpp{} code on top of Chroma~\cite{edwards:2004sx} from expressions of the form given in eqns.~(\ref{eq:3n_interp_op}) and~(\ref{eq:neutron_op}).
It consists of a chain of code written in FORM~\cite{vermaseren:2000nd}, Python, and \cpp.

The FORM script starts with high level, human readable expressions like (\ref{eq:3n_interp_op}) and generates all possible quark contractions.
The quark propagators are then collected into baryon blocks and split such that each sub-expression has a unique triple of blocks $B(x_1'|X_1) B(x_2'|X_2) B(x_3'|X_3)$ (not counting spin and colour indices).
All remaining terms are combined into $T$.

\section{Ensemble}\label{sec:ensemble}

We only present results for a single ensemble because computing three neutron correlators is computationally expensive despite the optimizations to the contraction code.
We generated Clover-Wilson fermions in 2+1 flavours with parameters modelled after the $96 \times 32^3$ CLS ensemble H107 presented in~\cite{bali:2016umi}, but with a larger spatial extent of $96 \times 48^3$.
The strange quark has physical mass $m_s = m_s^{\text{phys}}$ and the lattice spacing is $a = 0.085\unit{fm}$.
The pion mass as measured by RQCD on H107 is $m_\pi = 368\mev$ which gives $m_\pi L = 7.7$ on our lattices.
The analysis presented here uses $N_\text{cfg} = 175$ configurations and production is ongoing.
It is important to note that the preliminary analysis presented here uses point-to-point quark propagators, thus we expect cleaner signals in the final analysis.

\section{Results}

\subsection{Single Pion and Nucleon}

The code generator described above can handle smaller systems as well.
We used it to generate contraction codes for single pion and nucleon propagators as well as 2N correlators.
All results shown here have been computed for zero total momentum and total isospin $I$ and total spin $S$ as indicated in the figure titles.

\begin{figure}[h!]
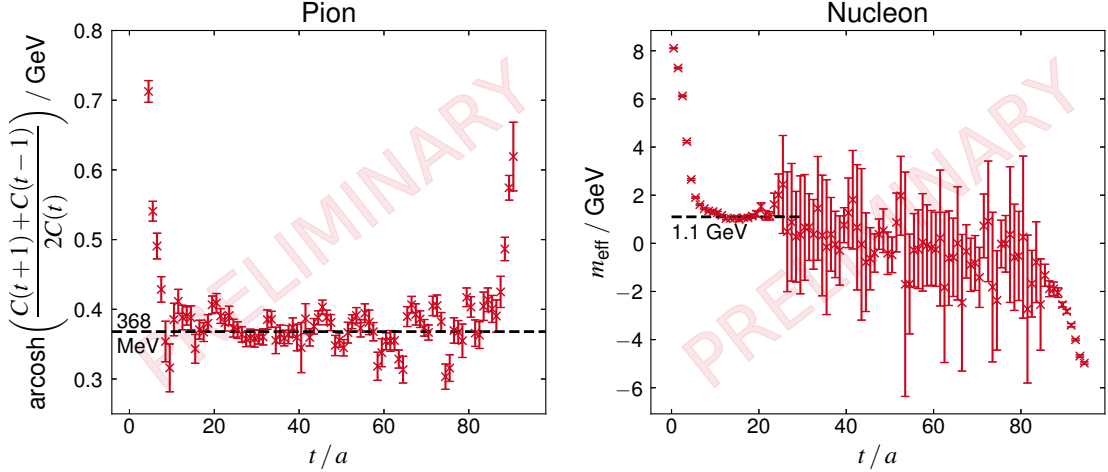

  \centering
  \resizebox{.48\textwidth}{!}{\input{./img/pi_meff.pgf}}
  \resizebox{.48\textwidth}{!}{\input{./img/n_meff.pgf}}
  \caption{Effective masses of single pion and nucleon $s$-wave propagators. The black lines are \textbf{not} fit results but rather indicate where the particle masses should be according to the measurements performed by~\cite{bali:2016umi}.\label{fig:pi_nucl_corr}}
\end{figure}

Figure~\ref{fig:pi_nucl_corr} shows effective masses for the $s$-wave propagators of pions and nucleons.
The effective mass of the pion is shown in symmetric form.
The black dashed lines in those figures are not fitted to the data.
In case of the pion, it shows the mass measured by the RQCD collaboration in~\cite{bali:2016umi}.
Our data seems to agree with this value but shows large fluctuations which should disappear once more statistics become available.
The $1.1\gev$ line in the single nucleon plot has been placed manually to roughly match data and agrees with the usually observed $m_N \approx m_\pi + 800\gev$~\cite{walker-loud:2014iea}.

\subsection{Two Nucleons}

Correlators for 2Ns can be cast into a form similar to~(\ref{eq:bblock_corr}) containing only two $B$s.
These expressions are much shorter than for three nucleons and it is feasible to treat all three relevant isospin channels.
Ultimately, this amounts to extending the tensor $T$ with isospin indices.
This formalism has been applied to the 2N system before.
Some $2N$ calculations can be found in~\cite{berkowitz:2018pts, orginos:2015aya, yamazaki:2015asa, berkowitz:2015eaa}.

Here, we follow the approach of~\cite{berkowitz:2015eaa} which places the quark propagators at spacially separated sources.
For the preliminary study presented here, we used only a single displacement with maximally separated sources $\Delta_\text{sources} = (\sfrac{L}{2},\, \sfrac{L}{2},\, \sfrac{L}{2})$.

\begin{figure}[h!]
  \centering
  \resizebox{.8\textwidth}{!}{\input{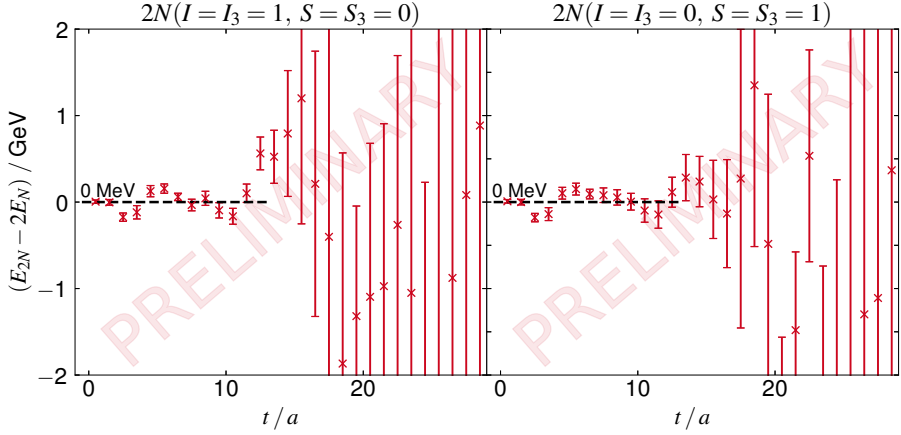}}
  \caption{Energy shifts of some $s$-wave two nucleon systems. The black lines are \textbf{not} fitted to the data. The $I=I_3=0$, $S=1$, $S_3=0$ channel is similar to the shown $S=S_3=1$ and has been omitted for clarity.\label{fig:two_n_corr}}
\end{figure}

Some results for $s$-wave two nucleon systems are shown in Figure~\ref{fig:two_n_corr}.
Results in the $I=I_3=0$, $S=1$, $S_3=0$ channel are essentially the same as the $S=S_3=1$ channel due to isospin symmetry.
The $S=S_3=0$ channel should be different however, which is reflected in the figure.
The combinations not shown here ($I=S=1$ and $I=S=0$) vanish for zero centre-of-mass momentum; we have confirmed this.

Again, the black lines are not fits to the data but have been placed to guide the eye.
Within the current statistics, the energy shifts are consistent with zero but do not show clear plateaus.

\subsection{Three Neutrons}

In the $3N$ systems, we focus on the novel three neutron correlators, i.e.\ maximal isospin.
The code generator is general enough to handle the other isospin channels as well however.
The other systems have already been discussed in~\cite{aoki:2012tk} and hence, for simplicity, we do not discuss them here.

All three quark sources have been placed at different lattice sites to avoid the Pauli principle.
In particular, the sources are at (omitting the time coordinate)
\begin{align*}
  x_1 = \big(0,\; 0,\; 0\big),\quad  x_2 = \big(\frac{L}{2},\; \frac{L}{2},\; \frac{L}{2}\big),\quad  x_3 = \big(\frac{L}{4},\; \frac{L}{4},\; \frac{L}{4}\big).
\end{align*}
The sinks are transformed into momentum space according to~\eqref{eq:mom_project} which produces relative $p$-waves with zero centre-of-mass momentum.

\begin{figure}[h!]
  \centering
  \resizebox{.8\textwidth}{!}{\input{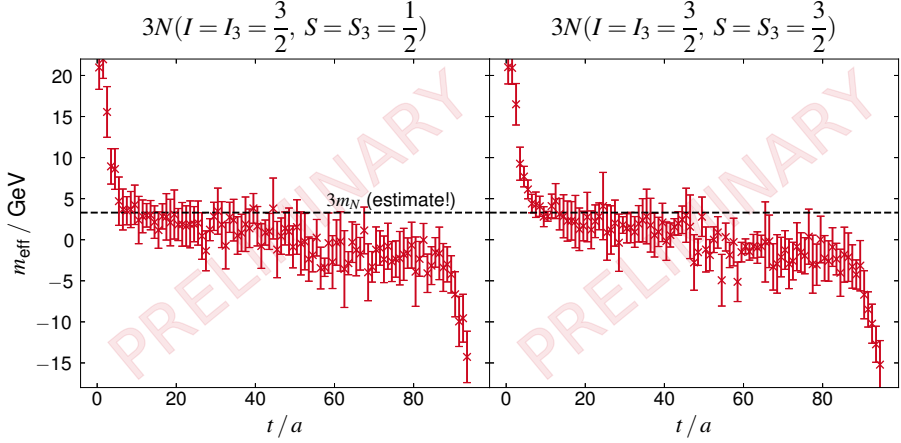}}
  \caption{Effective masses of three nucleon $p$-wave correlators. The black dashed lines are \textbf{not} fitted to the data but placed at $3 m_N$ as estimated in Figure~\ref{fig:pi_nucl_corr}.\label{fig:three_n_meff}}
\end{figure}

Figure~\ref{fig:three_n_meff} shows the resulting effective masses for $S=S_3=1/2$ and $S=S_3=3/2$.
The remaining channel, $S=3/2, S_3=1/2$, has been omitted because the data is qualitatively the same; as expected.
The statistical uncertainties are more than an order of magnitude larger than for two nucleons.
Contrary to the usual strong increase of noise for medium times, they are approximately constant across the whole range of $t$.
The plots show clear plateaus over long times which are consistent with the estimate of $3m_N$ in Figure~\ref{fig:pi_nucl_corr}.
This might be an artefact of the low statistical quality however.

\begin{figure}[h!]
  \centering
  \resizebox{.8\textwidth}{!}{\input{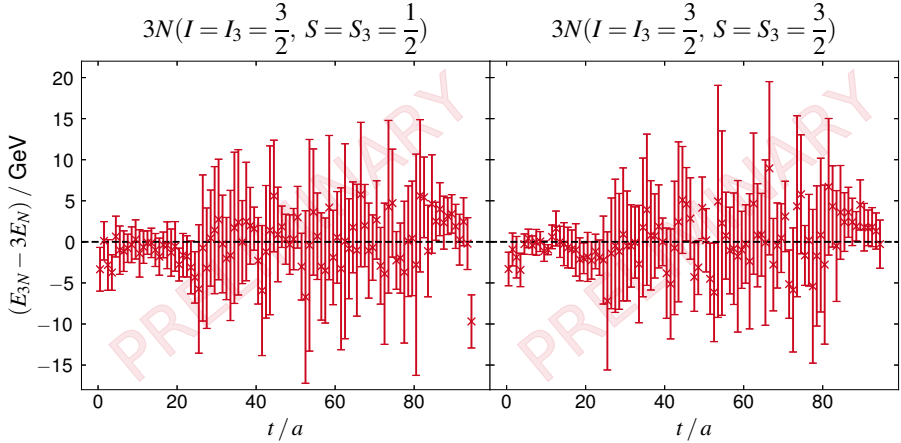}}
  \caption{Energy shifts of the three nucleon correlators. The raw effective masses are shown in Figure~\ref{fig:three_n_meff}. The black dashed line is a guide to indicate zero and \textbf{not} a fit to data.\label{fig:three_n_eshift}}
\end{figure}

The energy shifts with respect to $m_N$ are shown in Figure~\ref{fig:three_n_eshift}.
They exhibit an increase in statistical errors for larger times due to the single nucleon correlators.
For small times, there are semblances of plateaus consistent with zero.

\section{Conclusion and Outlook}

We have generalised the formalism of baryon blocks to three neutron interactions with fully displaced sources.
Based on this, we developed a source code generator to automatically produce and optimise contraction codes.
As a first application of this generator, we have looked at three neutron interactions on a newly produced gauge ensemble.
The preliminary results are promising and show the expected patterns.
Statistical uncertainties are too large however to perform a proper analysis for now.

Generation of configurations is ongoing which will allow us to gather more statistics.
In addition, we will perform more measurements on each configuration with smeared propagators and different source displacements to reduce both statistical noise and excited state contaminations.

\acknowledgments
The authors gratefully acknowledge the computing time granted by the JARA-HPC Vergabegremium and provided on the JARA-HPC supercomputers JURECA and JUQUEEN at Forschungszentrum J\"ulich.

\providecommand{\href}[2]{#2}\begingroup\raggedright\endgroup


\end{document}